\documentclass[a4paper,10pt]{article}

\usepackage[spacing]{microtype}
\usepackage[T1]{fontenc}
\usepackage{amsfonts,textcomp} 
\usepackage{amssymb,amsmath,amsthm}
\usepackage{textcase}
\usepackage{xcolor,hyperref}
\usepackage[round]{natbib}
\usepackage{units}
\usepackage{authblk}
\usepackage{fixfloats}
\usepackage{prolog,ocaml,abbrvs,refutils,figutils,nicefrac}

\lstMakeShortInline[language={[Samer]Caml},basicstyle=\rmfamily,identifierstyle=\itshape]| 
\lstMakeShortInline[language={[SWI]Prolog},basicstyle=\rmfamily,identifierstyle=\itshape]"

\def\nf(#1,#2){\nicefrac{#1}{#2}}

\definecolor{DarkRed}{rgb}{0.4,0.1,0}
\definecolor{MidRed}{rgb}{0.6,0,0}
\hypersetup{colorlinks,citecolor=black,urlcolor=MidRed,linkcolor=MidRed}

\usepackage{graphicx}
\graphicspath{{figs}}

\def\citepos#1{\citeauthor{#1}'s \citeyearpar{#1}\xspace}

\DeclareMathAlphabet{\mathcal}{OMS}{cmsy}{m}{n}

\def\TheAuthors{Samer Abdallah (samer@jukedeck.com)}
\def\TheAcknowledgments{The author would like to thank Jan Wielemaker, the
creator of SWI Prolog, for being very responsive in making enhancements to
and correcting problems in the implementation of delimited continuations 
in SWI Prolog, as well as being very helpful in explaining some of the
internal details of the implementation.}

\def\TheAuthors{Samer Abdallah (\texttt{samer@jukedeck.com})}
\def\TheInstitution{Jukedeck Ltd.}

\def\TheTitle{More declarative tabling in Prolog using multi-prompt delimited control}

\def\TheAbstract{%
Several Prolog implementations include a facility for \emph{tabling}, an alternative
resolution strategy which uses memoisation to avoid redundant duplication of
computations. Until relatively recently, tabling has required either low-level
support in the underlying Prolog engine, or extensive program transormation
\citep{GuzmanCarroHermenegildo2008}.
An alternative approach is to 
augment Prolog with low level support for continuation capturing control operators,
particularly \emph{delimited continuations}, which have been investigated in the 
field of functional programming and found to be capable of supporting a wide variety 
of computational effects within an otherwise declarative language. 

This technical report describes an implementation of tabling in SWI Prolog based
on delimited control operators for Prolog recently introduced by \cite{SchrijversDemoenDesouter2013}.
In comparison with a previous implementation of tabling for SWI Prolog using delimited
control \citep{DesouterVan-DoorenSchrijvers2015}, this approach, based on 
the functional memoising parser combinators of \cite{Johnson1995}, stays closer to
the declarative core of Prolog, requires less code, and is able to deliver solutions
from systems of tabled predicates incrementally (as opposed to finding all solutions
before delivering any to the rest of the program).

A collection of benchmarks shows that a small number of carefully targeted optimisations
yields performance within a factor of about 2 of the optimised version of Desouter 
et al.'s system currently included in SWI Prolog.
}

\title{\TheTitle}
\author{\TheAuthors}
\affil{\TheInstitution}

\begin{document}
	\maketitle
	\begin{abstract}\TheAbstract\end{abstract}
	
\section{Introduction and background}
\seclab{intro}

Tabling, or memoisation \citep{Michie1968} is a well known technique for speeding up computations
by saving and reusing the results of earlier subcomputations instead of repeating
them. In other contexts, this strategy is called `dynamic programming', and, when
used effectively, can reduce algorithms with exponential time complexity to polynomial
or linear complexity. 

In the field of logic programming, memoisation is referred to as \emph{tabling},
and usually includes the ability to handle recursive, non-deterministic predicates, including
so-called \emph{left-recursive} predicates that would otherwise result in non-termination using 
depth-first search strategies
such as SLD resolution, as provided, for example, in Prolog. Alternative strategies
such as OLDT resolution \citep{TamakiSato1986} and SLG resolution \citep{ChenWarren1993} have
been available for some time in Prologs such as XSB and YAP, but it was recognised
well before that the forms of deduction that these methods embody are closely related to
the efficient parsing algorithm invented by \cite{Earley1970}, the generalisation of which
was dubbed `Earley deduction' \citep{Warren1975,PereiraWarren1983,Porter1986}.

In the functional programming community, the problem of efficient parsing has been
addressed using memoising parser combinators \citep{Norvig1991,Leermakers1993}.
\cite{Abdallah2017a} gives an overview of this approach and focusses on the method of
\cite{Johnson1995}, which is particularly
interesting in that it handles left-recursive grammars using essentially the same 
mechanism as Earley's algorithm and, by extension, the OLDT and SLG resolution methods
used in tabled Prologs. Those methods rely on the \emph{suspension} of chains of deduction,
which can then be resumed under different conditions. Johnson's method (implemented in the
functional programming language Scheme) uses continuation
passing style \citep{SussmanSteele1975} to achieve the same effect, allowing threads of
computation to be suspended and resumed multiple times. Indeed, transformation of a program
into continuation passing style, or the use of continuation capturing control operators
such as |call-with-current-continuation| in Scheme, is a very powerful approach
to implementing many computational effects within a declarative language \citep{Filinski1999}.

Continuation passing style has been investigated as a basis for implementing Prolog itself
\citep{TarauDahl1994,Lindgren1994,Neumerkel1995} and for providing useful computational effects
from within in Prolog \citep{TarauDahl1994a,TarauDahl1998}. 
Continuations were also the basis of \citepos{GuzmanCarroHermenegildo2008} 
implementation of tabling, which required a fairly extensive source level program
transformation to make the continuations available.

Recently, continuation capturing control operators have been implemented in hProlog 
\citep{SchrijversDemoenDesouter2013}, and subsequently used to implement
tabling completely in Prolog, without any further low level support 
\citep{DesouterVan-DoorenSchrijvers2015}. This report describes a similar approach, first
translating \citepos{Johnson1995} functional memoisation into pure, declarative Prolog (in \secrf{functional}),
then (in \secrf{idiomatic}) making minimal changes to implement tabling as it is more usually seen in Prolog,
while retaining most of the declarativeness of the functional version, and finally (in \secrf{complexity}) analysing
the computation complexity of the system and making a small
number of targeted optimisations to yield acceptable performance. Concluding remarks are to be found
in \secrf{conclusions} and supplementary code in the appendices.

The system presented here was developed using SWI Prolog \citep{WielemakerSchrijversTriska2012},
which includes an implementation of delimited continuations
derived from \citepos{SchrijversDemoenDesouter2013} original implementation for hProlog.
We assume some familiarity with Prolog in general. In the program listings that follow, the code
has been rendered using typographic features not available in the real source; in particular
\verb|:-| is rendered as ":-", \verb|->| as "->", \verb|-->| as "-->", and variables with numeric suffixes
are rendered with subscripts. In addition, Prolog lambda expressions \citep{Neumerkel2009} are typeset, for example,
as "\X^Y^Goal", but coded as \verb|\X^Y^Goal|.

\section{Multi-prompt delimited control in Prolog}

Continuation passing style and continuation-based control operators have been well studied in the
context of lambda calculus and functional programming \citep{SussmanSteele1975}.
The idea of delimited continuations was arrived at independently by \cite{Felleisen1988} and
\cite{DanvyFilinski1990}. Multiple, first class, independently addressable prompts were introduced
by \cite{DyvbigJonesSabry2005} in their monadic Haskell implementation, which was recast
as an OCaml library to be used in `direct style' (i.e. using neither monadic nor explicit continuation 
passing styles) by \cite{Kiselyov2012}. The basic operation of multi-prompt delimited
control can be described in terms of the |new_prompt|, |push_prompt| and |shift| operators provided by
this library, which have the following OCaml types (in which Greek letters denote type variables):
\begin{ocamlet}
	new_prompt	: unit -> 'a prompt
	push_prompt	: 'a prompt -> (unit -> 'a) -> 'a
	shift        : 'a prompt -> (('b -> 'a) -> 'a) -> 'b
\end{ocamlet}
Given a prompt |p| with answer type |'a|, created using |new_prompt|, a thunk |f : unit -> 'a| is evaluated
in a context delimited by the prompt |p| by calling |push_prompt p f|, which plays the role of |reset| in 
\citepos{DanvyFilinski1990} framework.
Calls to |push_prompt|, using the same or differing prompts, can be nested arbitrarily.
If, inside |f|, there is a call to |shift p h|, the
continuation out to the nearest instance of |push_prompt p| is captured as a function |k : 'b -> 'a|, and passed to the handler
|h|, whose return value is then returned from the enclosing |push_prompt|. The captured continuation includes an implicit
enclosing |push_prompt p|, which hides any further calls to |shift p| it may contain, and means that, from the outside,
it looks like a pure function, which can be called zero, once, or many times. During a call to the handler |h|,
the original outer prompt stays in place, so that any calls to |shift p| made by the handler do not
escape the original delimited context. Other frameworks provide a different set of control operators, but
they are all closely related and expressible in terms of each other \citep{Shan2004}.

\citepos{SchrijversDemoenDesouter2013} implementation of delimited continuations for Prolog provides two
primitives, "reset/3" and "shift/1". If a Prolog goal is called as "reset(Goal, Ball, Cont)", with both
"Ball" and "Cont" unbound, then "Goal" is executed inside a delimited context. If it completes
successfully, then "Cont" is unified with "0".  If "shift/1" is called inside "Goal", then "Cont" is bound
to a callable representation of the continuation from that point out to the enclosing "reset", "Ball"
is bound to the argument of "shift/1", and control is passed to the code following the "reset".
The continuation "Cont" can be called subsequently, resuming the interrupted execution of the original
goal, but if that includes further calls to "shift/1", then it must be called inside a new
"reset", since, unlike in the functional version of |shift|, the captured continuation 
does \emph{not} re-establish the delimited context.

Calls to "reset/3" can be nested.  In this case, a "shift" is handled by the nearest
enclosing "reset", \emph{unless} that "reset" was called with a partially instantiated "Ball" parameter
that does not unify with the parameter to "shift/1". In this case, the "shift" travels outwards
until a "reset" with a unifiable "Ball" parameter is found. This mechanism is intended to allow different
computational effects to be implemented by nesting calls to "reset/3", each matching a different set of
"Ball" terms and each accompanied by a suitable effect handler to deal with the requested effect and then 
resume the computation by calling the continuation.

We will begin with a module "delimcc" that provides an alternative interface to delimited control that
uses a result term that can be tested by pattern matching on the head functor, instead of the `defaulty' 
"Cont" produced by "reset/3", which requires a test and a cut to be processed correctly.
It also 
and also avoids the non-logical behaviour with partially instantiated "Ball" terms by taking instead
an explicit first class prompt
as in \citepos{DyvbigJonesSabry2005} and \citepos{Kiselyov2012} libraries. 
\begin{prolog-framed}[name=delimcc]
  :- module(delimcc, [p_shift/2, p_reset/3, pr_shift/2, pr_reset/3]).

  :- meta_predicate p_reset(+,0,-).
  ~p_reset~(Prompt, Goal, Result) :-
    reset(Goal, Prompt-Signal, Cont),
    ( Cont == 0 -> Result = done
    ; Result = susp(Signal, Cont)
    ).

  ~p_shift~(Prompt, Signal) :- shift(Prompt-Signal).
\end{prolog-framed}
"Prompt" is a separate argument to both "p_reset/3" and "p_shift/3", and must be a
ground term. The result of the call to "p_reset/3" is either "done" to indicate no continuation
was captured, or "susp(Signal,Cont)" to indicate that "Signal" was sent to this prompt and
the captured continuation is "Cont".

With the aim of drawing a closer parallel between the Prolog and functional implementations of delimited
control, the module also contains "pr_reset/3" and "pr_shift/2", which build on "p_reset/3" and "p_shift/2".
Instead of the "Goal" argument of "p_reset/3", "pr_reset/3" accepts a callable term "Pred" representing a
unary predicate. This is called with a single unbound argument, as in "call(Pred, X)", and plays the role of the thunk
|f : unit -> 'a| taken by |push_prompt| in the OCaml library. Instead of the term "Signal" 
taken by "p_shift/2", "pr_shift/2" takes a term "H" representing a binary predicate and is called as
"call(H,K,X)", where "K" represents the continuation as a unary predicate 
and "X" is the unbound variable that will be the final answer returned by "pr_reset/3" in its third argument.
\begin{prolog-framed}[name=delimcc,firstnumber=15]
  :- use_module(lambdaki).

  :- meta_predicate pr_reset(+,1,-).
  ~pr_reset~(Prompt, Pred, Result) :-
     p_reset(Prompt, call(Pred, X), Status),
     pr_cont(Status, Prompt, X, Result).

  ~pr_cont~(done, _, X, X).
  ~pr_cont~(susp(H, Cont), Prompt, X, Result) :-
     pr_reset(Prompt, call(H, pr_reset(Prompt, \X^Cont)), Result).

  :- meta_predicate pr_shift(+,2).
  ~pr_shift~(Prompt, H) :- p_shift(Prompt, H).
\end{prolog-framed}
Note that, before being passed to the handler "H", the raw continuation "Cont" is wrapped in its
own delimited context
and expressed as a unary predicate by using a lambda term "\X^Cont" (see the supporting
module "lambdaki" in the appendix~\ref{sec:lambdaki}), which turns the 
output variable "X" into a parameter,
and introduces exactly the right copying semantics to allow the continuation to be
called multiple times without the interference that could be caused by the possible
binding of variables inside the original continuation term "Cont". Together these have the
effect of making the continuation look pure from the outside. Note also that the handler
is called inside a delimited context and so may use computational effects too, just as
in the functional version of |shift|.

\section{Memoisation functional style}
\seclab{functional}

\subsection{State handling using delimited control}
\label{sec:ccstate}

Stateful computations are sometimes handled in Prolog by using two extra arguments to pass the state in and out
of all predicates that need to manipulate it. This has the advantage of preserving a  
declarative reading of the program, without resorting to a procedural reading involving side effects, but done 
manually, it can become tedious and
error prone. The \emph{definite clause grammar}, or DCG syntax \cite{PereiraWarren1980}
recognised by most Prologs makes this much easier by hiding the extra arguments---they are inserted and
threaded through all goals automatically when source files containing DCG notation
are read. However, using DCG syntax in this way requires that all the code be lifted into DCG form
even if many parts do not need to manipulate the state. Delimited continuations allow statefulness
to be provided as a computation effect \emph{within} the delimited context, but to appear pure
(like a DCG) from outside the context. \cite{SchrijversDemoenDesouter2013} use state handling
as one of their examples, and we will use a similar approach here, but using the multi-prompt
control predicates introduced in the previous section to allow the safe nesting of state handling
contexts within each other or other effect handlers. The module "ccstate" provides,
within the context of "run_state/3", "app/2" to manipulate the state (using an arbitrary binary predicate
to model a state transition) as a computational effect.
\begin{prolog-framed}[name=ccstate]
  :- module(ccstate, [run_state/4, app/2]).
  :- use_module(library(delimcc)).

  :- meta_predicate app(+,2).
  ~app~(Pr,P)  :- p_shift(Pr,app(P)).

  :- meta_predicate run_state(+,0,?,?).
  ~run_state~(Pr, Goal, S1, S2) :- 
     p_reset(Pr, Goal, Status), 
     cont_state(Status, Pr, S1, S2).

  ~cont_state~(done, _, S, S).
  ~cont_state~(susp(app(P),Cont), Pr, S1, S3) :- 
    call(P, S1, S2), 
    run_state(Pr, Cont, S2, S3).
\end{prolog-framed}
The predicate "run_state/4" reifies this effect
as two extra arguments, that is, it makes stateful effects inside the given goal look like the result
of a pure predicate from the outside, with the last two arguments carrying the initial and final
states.

This interface provides access to a single `blob' of state, but it is quite 
straightforward to build on top of it an effect handler than instead provides an 
unbounded supply of references to mutable cells, which can then be manipulated independently:
\begin{prolog-framed}[name=ccstate]
  :- use_module(library(data/store)).

  :- meta_predicate run_ref(0), ref_app(+,2).
  ~run_ref~(Goal) :- store_new(S), run_state(ref, Goal, S, _).

  ~ref_new~(X,R)   :- app(ref, store_add(X,R)).
  ~ref_app~(R,P)   :- app(ref, store_apply(R,P)).
  ~ref_upd~(R,X,Y) :- app(ref, store_upd(R,X,Y)).
\end{prolog-framed}
The module "library(data/store)" (see appendix~\ref{sec:library}) is included in the SWI Prolog add-on package "genutils"
and provides a pure, declarative implementation of a reference-value store using 
immutable tree data structures.

\subsection{Memoised nondeterminism using lists}
\label{sec:ccmemo}

The facilities described so far are just enough to implement memoisation as a 
reasonably direct port of the functional, continuation-based approach implemented monadically
in OCaml by \cite{Abdallah2017a}, which was in turn based on \citepos{Johnson1995}
explicit CPS approach. It supports the memoisation of binary predicates (with one input
and one output), which may be recursive or indeed left-recursive. Nondeterminism is
invoked using "choose/2", implemented by the continuation handler "choose/4", and
finally reified as a list of alternative results. (Library "rbutils" is part of the "genutils"
add-on package mentioned above, and provides additional tools for working with the
red-black balanced binary trees implemented in the SWI Prolog built-in library "rbtrees".
See appendix~\ref{sec:library} for more details.)
\begin{prolog-framed}[name=ccmemo,numbers=left]
  :- module(ccmemo, [run_memo/2, choose/2, memo/2]).

  :- use_module(library(rbutils)).
  :- use_module(library(delimcc)).
  :- use_module(library(ccstate)).
  :- use_module(lambdaki).

  :- meta_predicate memo(2,-).
  ~memo~(P, ccmemo:mem_call(P,R)) :-
     rb_empty(T),
     ref_new(T,R).

  ~mem_call~(P,R,X,Y) :- pr_shift(memo, h_mem(P,R,X,Y)).
  ~choose~(Xs,X) :- pr_shift(memo, h_choose(Xs,X)).
  :- meta_predicate run_memo(1,-).
  ~run_memo~(P,Result) :- pr_reset(memo, to_list(P), Result).
  ~to_list~(P,[X]) :- call(P,X).

  ~h_choose~(Xs,X,K,Ys) :- foldl(call_append(\X^K),Xs,[],Ys).

  ~h_mem~(P,R,X,Y,K,Ans) :-
     Ky =\Y^K,
     ref_upd(R, T1, T2),
     (  rb_upd(X, entry(Ys,Conts), entry(Ys,[Ky|Conts]), T1, T2)
     -> rb_fold(fst_call_append(Ky), Ys, [], Ans)
     ;  rb_empty(EmptySet),
        rb_add(X, entry(EmptySet,[]), T1, T2),
        call(P,X,Y1),
        ref_app(R, rb_upd(X, entry(Ys,Conts), entry(Ys2,Conts))),
        (  rb_add(Y1, t, YS, Ys2)
        -> foldl(flip_call_append(Y1), [Ky|Conts], [], Ans)
        ;  Ans = [], Ys2 = Ys
        )
     ).

  ~fst_call_append~(Ky,Y-_,A1,A2) :- call_append(Ky,Y,A1,A2).
  ~flip_call_append~(Y,Ky,A1,A2) :- call_append(Ky,Y,A1,A2).
  ~call_append~(Ky,Y,A1,A2) :- call(Ky,Y,A), append(A1,A,A2).
\end{prolog-framed}
The context for running memoised nondeterministic computations is provided by "run_memo/2", which
installs a prompt named "memo" and calls the supplied predicate "P" using wrapper "to_list/2", which
returns the result of "P" in a (singleton) list. Nondeterminism is handled by "h_choose/4", which applies the
continuation "K" to each item in the supplied list of alternatives "Xs", concatenating the lists of results
produced by each application of the continuation. 

A memoised version of a binary predicate is 
prepared using "memo/2", which creates a new mutable reference containing the initially
empty memo table for that predicate. The memoised version is a call to "mem_call/4", 
which, invokes the continuation handler "h_mem/6". This works very much like 
\citepos{Abdallah2017a} functional version: if the memoised predicate has already been applied to
the input "X", then there will already be an entry in the memo table "T1" (line 25), in which case the
newly captured continuation, in the form of a two-argument lambda term "Ky", is added to the 
list of `consumer' continuations associated with "X", and then applied to all
the results produced by the base predicate "P" so far (line 26), collecting all the 
final results in the list "Ans". If this is the first time the memoised predicate has been applied to 
"X", then a new `producer' is initiated  by adding an entry to the memo table (line 28), calling the base
predicate "P", and, for each result "Y1", producing an empty answer list if "Y1" is 
already in the results table (line 33) or sending "Y1" to all of the continuations waiting for the
result of applying "P" to "X". The whole of the handler predicate "h_mem/6" is called inside a
"memo" prompt, so that any nondeterminism or memoisation effects triggered inside the call to "P" 
(line 29) result in the following lines of code being executed multiple times and the answers being collected
in the correct way.

As an example of how this framework can be used, the following illustrates a left-recursive
predicate "path/2" which implements the transitive closure of "edge/2".
\begin{prolog-framed}[numbers=left]
  :- use_module(ccmemo).

  ~path_memo~(Start, Ends) :-
    memo(path(Path), Path),
    run_memo(call(Path,Start), Ends).

  ~path~(Path, N1, N3) :- 
    choose([0,1],U), 
    ( U = 0 -> edge(N1,N3)
    ; U = 1 -> call(Path,N1,N2), edge(N2,N3)
    ).

  ~edge~(a,X) :- choose([b,c],X).
  ~edge~(b,d).
  ~edge~(c,d).
  ~edge~(d,X) :- choose([],X).
\end{prolog-framed}
The predicate "path_memo/2" must be called inside a context providing mutable references, \eg, as
"run_ref(path_memo(Start, Ends))".
A few observations on this code are in order. Firstly, "path/3" is written in \emph{open recursive} style;
that is, rather than call itself directly when recursion is required, it calls the first argument "Path", which
is assumed to represent the memoised recursive path predicate. The recursion is tied-up without using
an explicit fixed-point operator simply by memoising "path(Path)" on line 4 and unifying the result with "Path"
(results in a cyclic term). Secondly, all nondeterminism in the
program, both in "edge/2" predicate and "path/3", must be represented using "choose/2",
rather than using Prolog's built-in nondeterminism. This results in a rather non-idiomatic Prolog style, for
example, we cannot simply omit the clause for "edge(d,X)" on line 16 and rely on normal Prolog failure to indicate that there is no edge 
from "d"; we must use "choose([],_)" to express this fact. Thirdly, although not a problem in this example,
the restriction to memoising only binary `input-output' predicates does not fit well with the Prolog norm
of using arbitrary arity predicates without restrictions on which arguments may be considered
`inputs' or `outputs'. In short, we have not really implemented \emph{Prolog} style tabling, but rather
transplanted a functional idiom into Prolog. These deficiencies are remedied in the next section.

\section{Idiomatic Prolog tabling}
\label{sec:idiomatic}

\subsection{Tabling with non-backtrackable state}
\label{sec:ccnbstate}

The Prolog implementation of delimited continuations interacts unproblematically 
with Prolog's nondeterminism and backtracking: any choice points created inside a
"reset/3" are preserved on leaving the delimited context, and subsequent failure will
cause backtracking to a choice point, undoing of variable bindings, and re-execution of 
subsequent goals both inside and outside the delimited context. This means, for example,
the state handling provided by "run_state/4" interacts with nondeterminism just as
a DCG or other pure Prolog predicate would, with state transitions being undone on backtracking.

Tabling relies on sharing state \emph{across} alternative branches of computation.
The implementation in the previous section dealt with this by eschewing Prolog nondeterminism
entirely and using lists to accumulate alternative results.
If we wish to use Prolog's nondeterminism to represent these alternatives, then
the state of the tables must be managed in such a way that changes are not reversed on backtracking.

These two ways of combining of state with nondeterminism (usually referred to as \emph{backtrackable}
and \emph{non-backtrackable} state respectively) can be represented
in a functional setting using \emph{monad transformers} \citep{LiangHudakJones1995}.
Within that framework, the two computational effects can be
layered in either order, with the two resulting monadic types making clear the difference
between them. If |'s| is the type of the state and nondeterminism is represented as a list,
the type of a monadic computation yielding values of type |'a| is |'s -> ('a * 's) list| 
when layering state on top of nondeterminism, but
|'s -> 'a list * 's| with nondeterminism on top of state. The former corresponds to backtrackable 
state, since there is a state associated with each nondeterministic alternative, while the
latter corresponds to non-backtrackable state, since there is a single final state resulting from
the traversal process that produced the list of alternatives. The fact that the final state depends
on the traversal process does not compromise the purity of the computation, because the mapping
from initial state to results plus final state is still deterministic and free of side effects.

Using Prolog's nondeterminism to represent a collection of results, that is, using a predicate
that succeeds once for each result, the
requirements for declarativeness are more stringent: the model theoretic meaning of a predicate is
a \emph{set} of tuples, invariant with respect to the strategy used to find them.
In general, the use of non-backtrackable state means that the sequence of state transitions depends
on the order in which the proof tree of a goal is explored, so it follows that, if a pure logical
reading is to be maintained, the state
should not be exposed in the interface of a predicate that uses it internally.

SWI Prolog provides a number of ways to manage non-backtrackable state, including dynamic predicates
("assert/1" and "retract/1"), a `recorded database', and non-backtrackable
global variables ("nb_setval/2" and "nb_getval/2"). All three of these 
(necessarily) involve a departure from pure Prolog semantics.
Using "run_nb_state/4" (implementation given in appendix~\ref{app:ccnbstate}), the scope of this 
impurity and the associated dangers of unprotected global mutable state are limited to 
the introduction of one mutable cell accessed via a carefully controlled computational effect,
 with essentially the same interface as the backtrackable state effect provided by "run_state/4" in module
"ccstate" (\secrf{ccstate}). 

\subsection{Handling arbitrary predicate arity and mode}
\label{sec:arity-and-mode}

The next requirement for an idiomatic Prolog interface to tabling is the ability to
table predicates of any arity
and with any pattern of argument instantiation (usually referred to as \emph{mode}). This can be done
by maintaining a separate table for each distinct combination of input terms and output
variables encountered in calls to the predicate, that is, treating calls which are
\emph{variants} \citep{SterlingShapiro1994} of each other as forming an equivalence class.
For example, "foo(a,X,Y)" and "foo(a,U,V)" are variants of each other, but "foo(a,X,Y)",
"foo(b,X,Y)" and "foo(a,X,X)" are not. Variant calls will have the same set of solutions and
can share the same table. In the sequel, an equivalence class of calls which are all variants
of each other will be called a `variant class'.\footnote{It should be noted that manipulating terms on
the basis of what variables they contain involves stepping up out of the pure, declarative
fragment of Prolog in to pure Prolog's \emph{metalanguage}. The fact that Prolog serves as
its own metalanguage means that this distinction is easily overlooked.} The safest way to manipulate
variant classes without resorting to extra-logical operations
is to represent them as ground terms as soon as possible, which can be done
using "numbervars/3". This replaces each distinct variable with a term "`&VAR'(N)", where "N"
is an integer. Thus, in the examples above "foo(a,X,Y)" will be represented as "foo(a,`&VAR'(0),`&VAR'(1))",
while "foo(a,X,X)" will be represented as "foo(a,`&VAR'(0),`&VAR'(0))". 

Ground terms representing variant classes can be
used as keys to their associated tables in an associative map which is in turn stored in
the non-backtrackable mutable cell provided "run_nb_state/4", described in \secrf{ccnbstate}.
Because the key is derived from the predicate call, there is no need
for an explicit allocation step as there was in the functional version. Note also that the
stratification in table access between the binary predicate at the first level
and the input argument at the second is gone: each variant class combines predicate
and `input' arguments, and is a separate entity at the top level.

The `output' from calling a particular variant class consists of
the bindings of the variables in the call, which may be distributed in any
way across the parameters of the predicate, including possible partial instantiations
of arguments. These variables can be extracted using the standard Prolog extra-logical
operation "term_variables/2". Each solution of a tabled variant class can then be represented
as a list of the resulting values of these variables.

\subsection{Using Prolog nondeterminism}

The final requirement is to use Prolog's nondeterminism both at the level of the predicates
being tabled (instead of the awkward "choose/2" effect in the functional version)
and to represent nondeterminism outside the tabled environment (instead of reifying nondeterminism
as a list of alternatives). This will have the additional effect that the solutions of tabled
predicates can be explored incrementally, instead of being produced all at once.
In the module "cctable" below, this behaviour is effected in two places. Firstly, if a call is made
that already has a table entry (\ie a consumer call), instead of calling the consumer continuation
for each solution already found and aggregating the final answers, line 27 extracts a solution
from the solution set \emph{nondeterministically}, leaving a choice point, and then calls
the consumer continuation. Any subsequent failure will result in backtracking to this choice
point and application of the continuation to the next solution in the set.
Secondly, in "producer/4" (line 37), "call(Generate, Y)" will be nondeterministic, and each time it succeeds,
previously seen solutions can be ignored simply by failing in the attempt to add the value to the 
solution set (line 44),
while previously unseen solutions are sent to the registered continuations (including the producer's
continuation) by calling them \emph{disjunctively}, rather than conjunctively (line 40). It is significant that
the producer's continuation "KP" is called \emph{first}: this means that from the outside,
the tabled predicate produces answers lazily, before all solutions have been found.

\begin{prolog-framed}[name=cctable,numbers=left]
  :- module(cctable, [run_tabled/1, cctabled/1]).

  :- use_module(library(delimcc), [p_reset/3, p_shift/2]).
  :- use_module(library(ccnbstate), [run_nb_state/4, app/4]).
  :- use_module(library(rbutils)).
  :- use_module(lambdaki).

  :- meta_predicate cctabled(0).
  ~cctabled~(Work) :- p_shift(tab, Work).

  :- meta_predicate run_tabled(0).
  ~run_tabled~(Goal) :-
     term_variables(Goal, Ans), 
     rb_empty(Empty),
     run_nb_state(state, run_tab(Goal, Ans), Empty, _).

  ~run_tab~(Goal, Ans) :-
     p_reset(tab, Goal, Status),
     cont_tab(Status, Ans).

  ~cont_tab~(done, _).
  ~cont_tab~(susp(Work, Cont), Ans) :-
     term_variables(Work,Y), 
     K = \Y^Ans^Cont,
     numbervars_copy(Work, VC),
     app(state, new_cont(VC, K, A)),
     (  A = solns(Ys) -> rb_in(Y, _, Ys), run_tab(Cont, Ans)
     ;  A = producer -> run_tab(producer(VC, \Y^Work, K, Ans), Ans)
     ).

  ~new_cont~(VC, K, solns(Ys), S1, S2) :- 
     rb_upd(VC, tab(Ys,Ks), tab(Ys,[K|Ks]), S1, S2).
  ~new_cont~(VC, K, producer, S1, S2) :-
     rb_empty(Ys), 
     rb_add(VC, tab(Ys,[K]), S1, S2).

  ~producer~(VC, Generate, KP, Ans) :-
     call(Generate, Y),
     app(state, new_soln(VC, Y, Ks)),
     member(K,[KP|Ks]), call(K,Y,Ans).

  ~new_soln~(VC, Y, Ks, S1, S2) :-
     rb_upd(VC, tab(Ys1, Ks), tab(Ys2, Ks), S1, S2),
     rb_add(Y, t, Ys1, Ys2).

  ~numbervars_copy~(Work, VC) :-
     copy_term_nat(Work, VC),
     numbervars(VC, 0, _).
\end{prolog-framed}
A few other comments on this code are in order. 
The state of the system is represented as an associative map (using red-black trees)
of variant classes to table entries, each of which contains a set of solutions
(also represented as a red-black tree) and a list of consumer continuations. Each instance
of the state is an \emph{immutable} data structure: only the cell managed by
"run_nb_state/4" is mutable, and its contents are replaced completely on each state
transition. These transitions are managed using the predicates "new_cont/5"
(to decide what to do with the newly captured continuation on each call to a tabled
predicate) and "new_soln/4" (to decide what to do with a new solution from the 
`worker' goal "Work". These two predicates are themselves pure, so that the effectful
component of the state transition is completely encapsulated by the "app/2" operator.

The variable "Ans" is used 
to represent the list of `outputs' from the top-level goal initially passed
to "run_tabled/1"; this list is extracted from the goal using "term_variables/2".
Similarly, "Y" represents the `outputs' of a given tabled variant class as a list
of variables, which are eventually bound when solutions are found.

The continuation "Cont" captured when
calling a tabled predicate is a callable term, possibly containing variables shared
with the worker goal, representing `inputs' which are expected to be 
provided by the tabled predicate, and `outputs'
which will be bound when the continuation is called. This continuation is stored
in the table entry for a variant class in the form of a lambda term
"\Y^Ans^Cont" (line 24) so that it can be called as a binary predicate and called multiple times
without binding the variables in the lambda term. Similarly, the worker goal 
is passed to "producer/4" in the form of a lambda term "\Y^Work" (line 28),
so that it can be called without binding any of these variables.

\subsection{Program transformation}
While the module "cctable" can be used as-is, by calling tabled goals using the
meta-predicate "cctabled/1", the usual approach to tabling in Prolog is to support a
declaration that a certain predicate should be tabled, such that it can then be
called without any further decoration. This can be achieved using a very shallow
program transformation, similar to that described by \cite{DesouterVan-DoorenSchrijvers2015}.
The implementation is given in the appendix, but result is that the example given 
in \secrf{ccmemo} can be written (now using DCG notation for "path/2") as:
\begin{prolog-framed}
  :- use_module(cctable).
  :- use_module(ccmacros).

  :- table path/2.
  ~path~ --> edge; path, edge.

  ~edge~(a,b).  ~edge~(a,c).  ~edge~(b,d).  ~edge~(c,d).
\end{prolog-framed}
The "path/2" predicate is transformed via macro expansion on loading to
\begin{prolog}
  ~path~(N1, N2)     :- cctabled(`path#'(N1, N2)).
  `~path#~'(N1, N3) :- edge(N1, N3); path(N1, N2), edge(N2, N3).
\end{prolog}
As well as the DCG translation, the original predicate is renamed by
adding "`#'" to the name, while "path/2" is redirected to a tabled
call of the renamed predicate.

\section{Computational complexity and performance}
\seclab{complexity}

The system presented above achieves the aim of supporting
idiomatic Prolog tabling while consisting almost entirely of pure, declarative Prolog.
Unfortunately, its performance is terrible. This is because maintaining non-backtrackable
state in Prolog requires a certain amount of term copying: not only is this relatively
expensive in itself, it also tends to create more work for the garbage collector. 
Because the state is held in a \emph{single} mutable variable, the \emph{entire} state is
copied every on every transition, even if only a small component of the
state has been changed. This creates a tension between the desire for logical purity,
which drives us to minimise the number of points of mutation, and the desire for
efficiency, which drives us to \emph{factorise} the state into several
pieces, so that one component of the state can be modified without
incurring the cost of copying the rest. At this point, development becomes a 
question of pragmatic balance, attempting to achieve acceptable performance with
a minimal departure from logical purity, which is, perhaps, a little unsatisfying from a 
theoretical point of view, but seems to be unavoidable within the framework
of current Prolog systems. 

\subsection{Complexity analysis}
\seclab{complexity-metrics}

We can begin to quantify the costs of various copying operations as follows.
Given an application involving tabled predicates, let $N_p$ be the number
of `producer' calls; that is, the number of distinct variant classes in use.
Let $R_c$ be the average number of `consumer' calls per producer; that is, the
number of subsequent calls with a given variant class after the initial
one. Finally, let $R_s$ be the average number of solutions per producer.
It is clear that the tabling system must call each continuation for each solution
for each variant class, so that a minimal complexity for a computation that
processes every solution of every tabled call is $O(N_p R_c R_s)$. However,
decisions about to represent the state of the system can incur additional copying costs,
some of which are quadratic in one or more of these metrics. For example, if a piece of
state contains one item for each producer, then copying this state to add a new
producer results in an $O(N_p^2 K)$ cost overall, where $K$ is the size
of the data structure associated with each producer, which may be $O(1)$, $O(R_c)$,
$O(R_s)$ or $O(R_c R_s)$ depending on the representation chosen.
The following table summarises this and other costs that may apply:

\begin{center}
\begin{tabular}{ll}
  \emph{operation} & \emph{cost} \\
  \hline
  1. copying each continuation before calling    & $N_p R_c R_s$ \\
  2. copying each worker to get variant class    & $N_p R_c$ \\
  3. copying all producers to add a new producer & $N_p^2 K $ \\
  4. copying all solutions to add a consumer     & $N_p R_c R_s$ \\
  5. copying all continuations to add a solution & $N_p R_c R_s$ \\
  6. copying all continuations to add a consumer & $N_p R_c^2$ \\
  7. copying all solutions to add a solution     & $N_p R_s^2$ \\
\end{tabular}
\end{center}
Of these, the first cannot be avoided, since a single continuation may need to
be called multiple times with different `inputs' in the course of a computation.
The others result from representational choices; for example, (3) applies when
a single term contains information about all producers, which is the case
in the implementation of \secrf{idiomatic}, with $K = R_c R_s$; (6) applies
when all the continuations for a particular producer are stored in a single
data structure, and similarly (7) when all the solutions from a producer are
stored in one term. Indeed, the implementation of \secrf{idiomatic} suffers from
all these costs, in particular the quadratic ones, which explains why it is so slow
on all but the smallest problems.

There are many strategies for reducing or eliminating these copying costs,
several of which are explored in the code repository accompanying this report.%
\footnote{\url{https://github.com/samer--/cctable}}
Although a complete description of these is beyond the scope of this report, a few
strategies will be discussed below.

Firstly, it is clear from the table of copying costs that for asymptotic 
performance of $O(N_p R_c R_s)$ it is most important to avoid costs that are quadratic 
in $N_p$, $R_c$ or $R_s$, as these will eventually dominate in large problems.
The implementation "cctable(db)" does this by using dynamic predicates 
to achieve a complete factorisation of the state of
the system. Although the overheads associated with using dynamic predicates are larger
than those for non-backtrackable global variables, the asymptotic complexity means
that performance is still better than some of the implementations (not shown in the table)
using non-backtrackable global variables to carry less completely factorised state.

Another way to avoid a quadratic cost when adding solutions is to use the SWI Prolog
"tries" library, which allows a single immutable reference to point to a mutable
associative map. This was introduced to support the tabling library currently included
in SWI Prolog, but is sufficiently general to be used in the implementation "cctable(trie)", 
for both the map of variant classes to table entries and the set of solutions for each
variant class.
Tries are not a suitable data structure for the collection of continuations associated
with each producer, as it is not appropriate to treat them as a set or test them for equality.
Hence, to avoid quadratic costs here, in "cctable(trie)", a system of references to 
growable lists is introduced, built using
the low-level "nb_linkval/2" operation to avoid copying the continuations already in the
list, in combination with "ccnbref", a module providing a delimited context for safely managing 
non-backtrackable mutable storage cells (see appendix~\ref{app:ccnbref}).

While developing approaches to managing the collection of continuations efficiently,
it became clear that size of the continuation terms can be a significant factor, 
most obviously in the Fibonacci benchmark, where the total size
of the continuations was found to grow quadratically with the size of the problem.
Discussions with Jan Wielemaker, the main author of SWI Prolog, lead to the conclusion
that there were two causes of this. Firstly (and most easily dealt with) the inclusion
of the producer's continuation ("KP") as a parameter to "producer/4" meant that continuations
were being included as values in the stack frames as part of other subsequently captured continuations.
This can be avoided by including the producer's continuation in the data structure used
to store the consumers' continuations, rather than passing it as a parameter. As long
as producer's continuation is kept at the head of the list, the 
`incrementality' of the system (solutions being delivered as they are found)
is preserved.

The second cause was that the process of 
repeatedly capturing and reactivating continuations along a single conjunctive computational path
resulted in continuations which gradually accumulated non-operative stack frames.
This lead Jan Wielemaker to introduce optimisations in the low level implementation of 
"shift/1", similar in spirit to \emph{tail call}
optimisations, to remove these stack frames.
This solved the continuation growth problem 
completely, resulting in approximately linear time complexity on the Fibonacci benchmark using
the implementation "cctable(trie)/kp".
An equivalent modification was also applied to the dynamic database version 
to get the implementation "cctable(db)/kp".

\subsection{An improved implementation}

The implementation "cctable(trie)/kp" is given below. Continuations are represented using
the "k/3" functor instead of lambda terms, since a bulk copy (in "lref_get/2") is faster
than copying them individually when calling lambda terms. 
\begin{prolog-framed}[name=cctable_trie_kp,numbers=left]
  :- use_module(library(delimcc), [p_reset/3, p_shift/2]).
  :- use_module(ccnbref, [run_nb_ref/1, nbref_new/2]).
  :- use_module(lambdaki).
        
  :- meta_predicate cctabled(0). 
  ~cctabled~(Head) :- p_shift(tab, Head).

  :- meta_predicate run_tabled(0). 
  ~run_tabled~(Goal) :-
     term_variables(Goal, Ans), trie_new(Trie),
     run_nb_ref(run_tab(Goal, Trie, Ans)),

  ~run_tab~(Goal, Trie, Ans) :-
     p_reset(tab, Goal, Status),
     cont_tab(Status, Trie, Ans).

  ~cont_tab~(done, _, _, _).
  ~cont_tab~(susp(Head, Cont), Trie, Ans) :-  
     term_variables(Head,Y), K = k(Y,Ans,Cont),
     (  trie_lookup(Trie, Head, tab(Solns,Conts))
     -> lref_add(Conts, K),
        trie_gen(Solns, Y, _),
        run_tab(Cont, Trie, Ans)
     ;  lref_new(K, Conts), trie_new(Solns),
        trie_insert(Trie, Head, tab(Solns,Conts)),
        run_tab(producer(\Y^Head, Conts, Solns, Ans), Trie, Ans)
     ).

  ~producer~(Generate, Conts, Solns, Ans) :-
     call(Generate, Y),
     trie_insert(Solns, Y, t),
     lref_get(Conts,Ks), 
     member(k(Y,Ans,Cont),Ks), call(Cont).

  ~lref_new~(K0, Ref) :- nbref_new([K0], Ref).
  ~lref_get~(Ref, Xs) :- nb_getval(Ref, Ys), copy_term(Ys,Xs).
  ~lref_add~(Ref, K) :- 
    nb_getval(Ref, [K0|Ks]), duplicate_term(K,K1), 
    nb_linkval(Ref, [K0,K1|Ks]).
\end{prolog-framed}

\subsection{Benchmarks}

Several of the implementation variations described above were tested on a selection
of the benchmarks that were used by \cite{DesouterVan-DoorenSchrijvers2015} and
compared with alternative tabling implementations. 
Desouter et al.'s system was tested in two forms: the original all-Prolog
library, referred to below as "desouter(pl)", is a port of the original from hProlog to SWI Prolog\footnote{%
Available at \url{https://github.com/JanWielemaker/tabling_library}}, while
"desouter(plc)" is an optimised version, with some components re-implemented
in C for greater speed. In addition, the tabling implementations in two other Prologs,
YAP and B-Prolog were also tested, though it should be noted that these are both
much faster Prologs in general and that in both tabling is implemented at a low level.
Also, YAP does not support arbitrary precision arithmetic, and so cannot run the Fibonacci benchmark
with the parameters indicated.
The benchmarks themselves were derived from code available online.\footnote{\url{https://github.com/JanWielemaker/tabling_benchmarks}}

Each of these benchmarks results in a certain pattern of tabled calls, which can be characterised
as in \secrf{complexity-metrics} in terms of the number of producers, consumers and solutions, as follows:
\begin{center}
  \begin{tabular}{@{}l@{\qquad}rrrrr@{}}
 & $N_p$ & $N_c$ & $N_s$ & $R_c$ & $R_s$\\
\hline{"fib(1000)"} & 1001 & 998 & 1001 & 1.0 & 1.0\\
{"fib(2000)"} & 2001 & 1998 & 2001 & 1.0 & 1.0\\
{"nrev(500)"} & 501 & 0 & 501 & 0.0 & 1.0\\
{"nrev(1000)"} & 1001 & 0 & 1001 & 0.0 & 1.0\\
{"shuttle(2000)"} & 1 & 2 & 4001 & 2.0 & 4001.0\\
{"shuttle(5000)"} & 1 & 2 & 10001 & 2.0 & 10001.0\\
{"ping_pong(10000)"} & 2 & 2 & 20002 & 1.0 & 10001.0\\
{"path_dfst(50)"} & 50 & 2402 & 2401 & 48.0 & 48.0\\
{"path_dfst(100)"} & 100 & 9802 & 9801 & 98.0 & 98.0\\
{"path_dfst_loop(50)"} & 50 & 4803 & 4802 & 96.1 & 96.0\\
{"recognise(20000)"} & 2 & 2 & 20000 & 1.0 & 10000.0\\
{"pyramid(500)"} & 500 & 995 & 186751 & 2.0 & 373.5\\
{"test_joins"} & 1 & 4 & 371293 & 4.0 & 371293.0\\
{"monoidal"} & 35 & 42195 & 2664 & 1205.6 & 76.1\\
\hline\end{tabular}

\end{center}
Notably, the na\"ive reverse benchmark "nrev(_)" does not involve any consumer calls: each variant class
is called just once, producing one answer, and so this benchmark does not benefit from tabling at all.
Thus, the timing results give an indication of the minimum overhead introduced by the tabling system.
The results are shown in \tabrf{timings}, and a subset of the results illustrated graphically in
\figrf{speed}. The benchmarks were run on a 2012 MacBook Pro with a \unit[2.5]{GHz} Core i5 processor and 
\unit[8]{GB} of memory.

\begin{table}
\small%
\begin{center}
\newcommand\plh[1]{\lstinline[language=Prolog]{#1}}
\begin{tabular}{@{}lrrrrrrrr@{}}
 & \rotatebox{90}{\plh{yap}} & \rotatebox{90}{\plh{bprolog}} & \rotatebox{90}{\plh{desouter(plc)}} & \rotatebox{90}{\plh{desouter(pl)}} & \rotatebox{90}{\plh{cctable(db)}} & \rotatebox{90}{\plh{cctable(db)/kp}} & \rotatebox{90}{\plh{cctable(trie)}} & \rotatebox{90}{\plh{cctable(trie)/kp}}\\
\hline{"monoidal"} & \underline{1.3 s} & \textbf{881} & 2.2 s & 31 s & 146 s & 133 s & 3.8 s & 3.9 s\\
{"test_joins"} & \textbf{674} & 1.6 s & \underline{1.2 s} & 12 s & $\bot$ & $\bot$ & 2.2 s & 2.7 s\\
{"fib(1000)"} & -- & \textbf{2} & \underline{10} & 110 & 1.1 s & 110 & 240 & 20\\
{"fib(2000)"} & -- & \textbf{8} & \underline{30} & 230 & 4.6 s & 380 & 940 & 40\\
{"nrev(500)"} & \textbf{17} & 64 & 120 & 1.5 s & \underline{60} & 100 & 80 & 90\\
{"nrev(1000)"} & \textbf{65} & 253 & 490 & 5.5 s & \underline{250} & 380 & 310 & 330\\
{"path_dfst(50)"} & \textbf{3} & \underline{5} & 30 & 500 & 2.0 s & 2.0 s & 70 & 50\\
{"path_dfst(100)"} & \textbf{21} & \underline{39} & 260 & 3.9 s & 65 s & 84 s & 390 & 370\\
{"path_dfst_loop(50)"} & \textbf{15} & \underline{32} & 190 & 2.6 s & 23 s & 25 s & 590 & 330\\
{"ping_pong(10000)"} & \textbf{6} & 3.5 s & \underline{40} & 850 & 7.5 s & 7.9 s & 60 & 60\\
{"pyramid(500)"} & \underline{53} & \textbf{38} & 250 & 6.5 s & $\bot$ & $\bot$ & 380 & 570\\
{"recognise(20000)"} & \textbf{15} & 66 s & \underline{70} & 920 & 14 s & 14 s & 170 & 120\\
{"shuttle(5000)"} & \textbf{6} & 1.4 s & \underline{30} & 380 & 2.3 s & 2.2 s & 80 & 80\\
{"shuttle(10000)"} & \textbf{10} & 6.0 s & \underline{50} & 790 & 9.7 s & 9.9 s & 190 & 170\\
\hline\end{tabular}

\end{center}
\caption{Execution times for a variety of benchmarks using several tabling implementations.
Times are milliseconds unless explicitly stated in seconds. In each row, the best time
is typeset in boldface and the second best time is underlined. YAP does not include
support for arbitrary precision integer arithmetic and so was not able to run
the Fibonacci benchmark at the given sizes. A $\bot$ indicates that the benchmark
did not complete within the allowed time of \unit[240]{s}. \tablab{timings}}
\end{table}

\begin{figure}
  \begin{center}
    \hspace*{-1.2em}%
    \begin{tabular}{@{}c@{\hspace{-2em}}c@{}}
      \colfig[0.58]{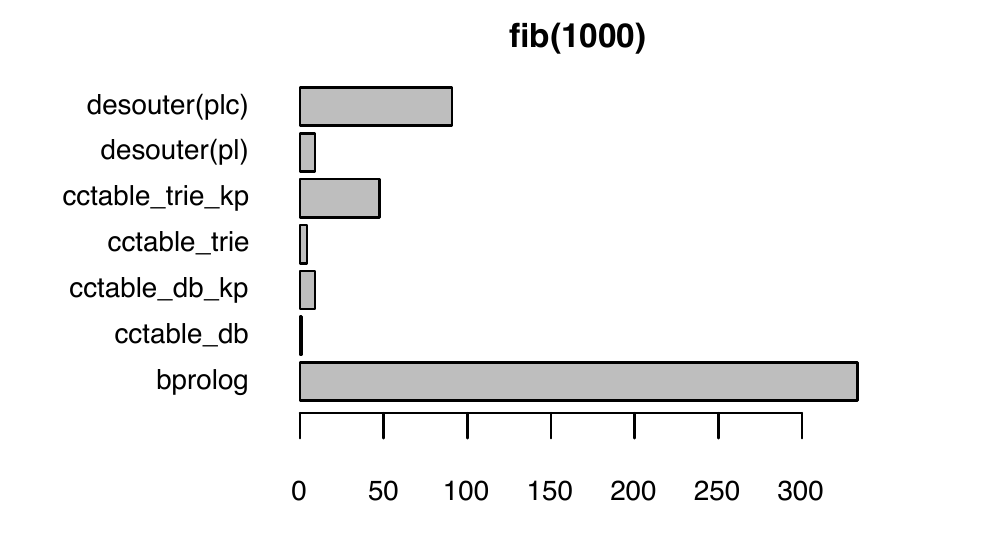}
    & \colfig[0.58]{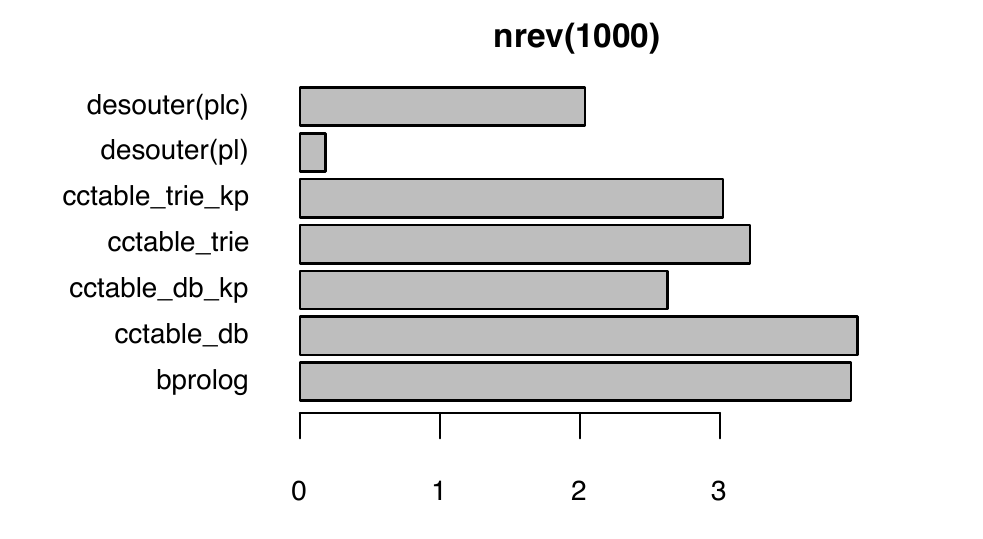}
    \\
      \colfig[0.58]{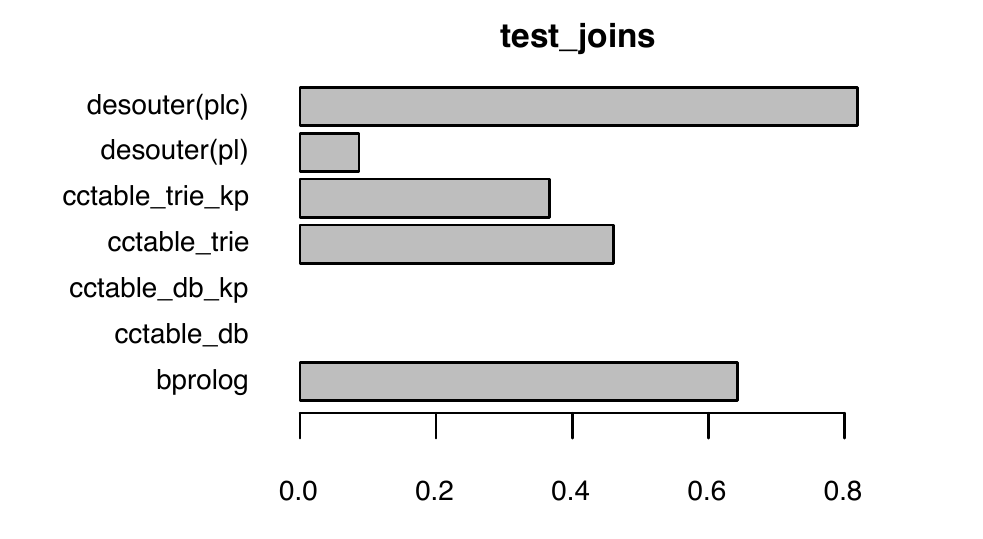}
    & \colfig[0.58]{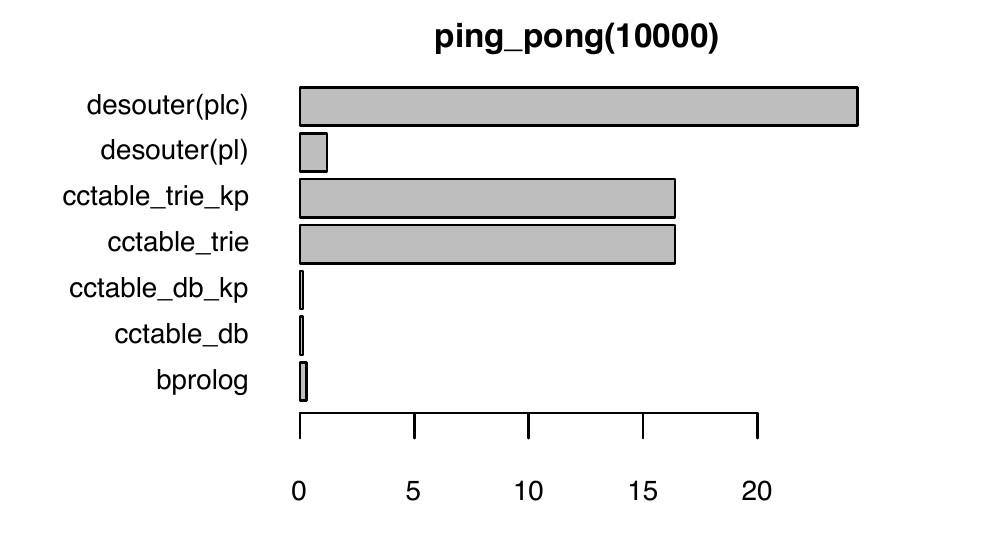}
    \\
      \colfig[0.58]{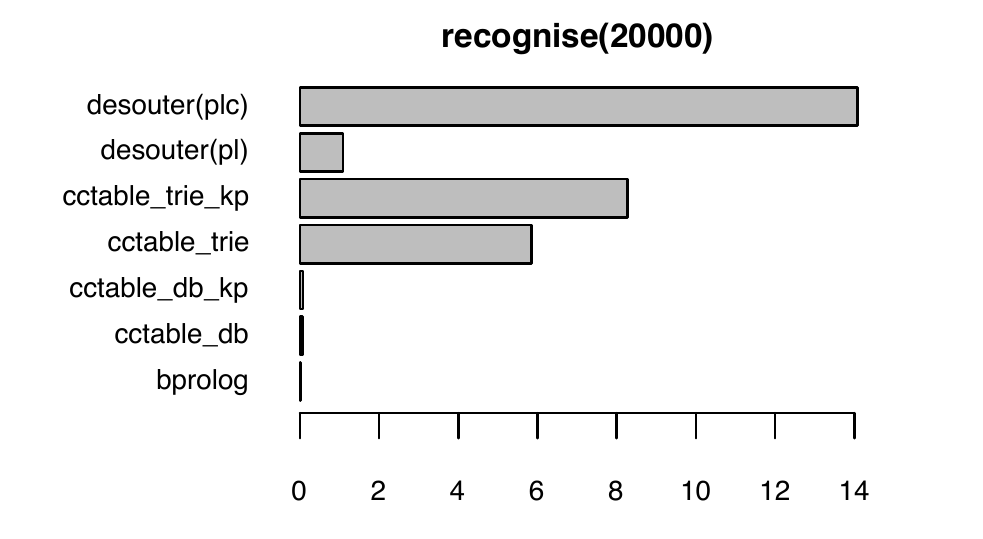}
    & \colfig[0.58]{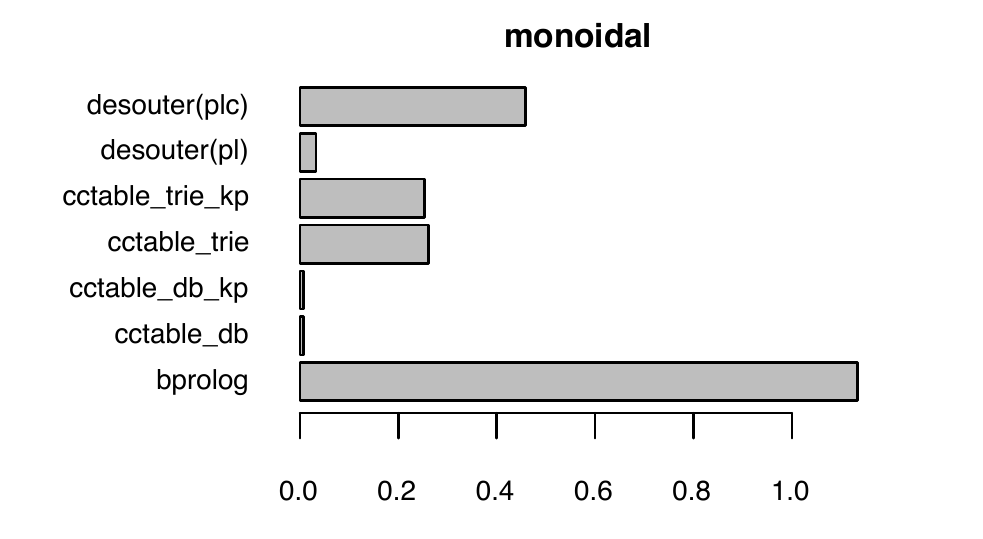}
  \end{tabular}
  \end{center}
  \caption{Performance of several tabling implementations on a subset of benchmarks. The horizontal
  axis is the reciprocal of the execution time in seconds, and so longer bars represent faster
  performance. The timings for YAP have been excluded as they are in almost all cases the fastest.
  \figlab{speed}}
\end{figure}

Overall, YAP is the fastest system, with efficient an implementation of SLG resolution as well as 
a fast underlying Prolog.\footnote{%
\cite{DesouterVan-DoorenSchrijvers2015} reported some of the execution times for YAP
at 0 ms. The code for running these
benchmarks measured the time to produce the \emph{first} solution produced. YAP's
tabling algorithm, like the one presented here, is able to produce solutions
as they are found, whereas Desouter et al.'s algorithm (both implementations) finds
all solutions before producing the first one. Hence the comparison is a bit misleading,
and so to produce the benchmarks here, the code was modified to measure the
time to find all solutions.}

B-Prolog uses a different tabling algorithm,
\emph{linear tabling} \citep{ShenYuanYou2001}, with quite different characteristics. It performs 
very well on many benchmarks but relatively poorly on those which involve a lot of non-determinism.

Of the SWI Prolog-based implementations, the optimised version of Desouter et al.'s library currently
included as standard in SWI is the fastest. Compared with this, The implementation "cctable(trie)/kp" 
developed here is quite competitive, running about half as fast in most benchmarks. Although
it shares the optimise trie implementation, it does not benefit from the optimised `worklist'
data structure, which stores and reactivates continuations more efficiently than is possible
from Prolog code.

It is clear from the results that the steps taken to avoid continuation
growth (the "_/kp" versions) are effective, especially on the Fibonacci benchmark, while having
a small negative impact on those benchmarks that did not suffer from excessive continuation
growth, with "pyramid(500)" suffering the most. It is also clear from the "nrev(_)" results
that the "cctable(db)" versions, using dynamic predicates, have the smallest overhead when
tabling turns out not be useful. However, these two implementations are not
competetive on several of the other benchmarks.

\section{Discussion and conclusions}
\seclab{conclusions}

The implementations presented in this report were largely derived by analogy from the OCaml
version described by \cite{Abdallah2017a}, but as they are built on the same set of delimited
control operators as the tabling library of \cite{DesouterVan-DoorenSchrijvers2015}, they all
share many common features: non-backtrackable mutable data structures for mapping variant calls
to table data; structures for collecting solutions; structures for collecting continuations and
reactivating them when solutions become available, and so on.

However, there are several significant differences.
The control structure and scheduling mechanism is simpler and more declarative, with less
reliance on side effects. There is much less code: Desouter
et al quote a figure of 244 lines of code, excluding the implementation of tries,
whereas the main tabling code here is about 40 lines of code, plus about 20 for program transformation
and 20 or so in general purpose libraries.
In use, it delivers solutions incrementally, rather than having to collect them all before 
delivering the first one, which could amount to a significant saving in execution time
in certain applications.

It was found while profiling the initial and most declarative implementation that the
largest contributions to execution time came from term copying and garbage collection.
This is an unavoidable factor when using non-backtrackable state in Prolog, and tends to force
the introduction of more separate pieces of mutable state in order to get acceptable
performance. However, it is still possible to use delimited control operators, with a relatively
low cost, to mange these pieces of mutable state with a minimal impact on the declarativeness
of the rest of the program.


The fastest implementation presented here, "cctable(trie)/kp", is within a factor of about 
2 of the tabling system currently included in SWI Prolog, and so with a little more 
optimisation, could become a viable alternative, bringing the benefits of incrementality,
and greater code simplicity and declarativeness.

\bigskip
\noindent
\textbf{Acknowledgements}\\[0.5em]
\TheAcknowledgments

\appendix
\section{Supporting code}

Modules "lambdaki", "ccstate", "delimcc" and "rbutils" can be found in the SWI Prolog
add-on package "genutils", which can be installed in SWI Prolog with the command
"pack_install(genutils)". The rest of the code presented in this report can be
found (in some cases in a slightly expanded and generalised form) 
at \url{https://github.com/samer--/cctable}.

\subsection{Lambda terms}
\label{sec:lambdaki}

This module is a minimal implementation of lambda terms, which are callable terms representing
predicates taking one or more arguments, and can be useful for high-order programming,
reducing the need to define trivial auxiliary predicates. It is a simplified version of 
\citepos{Neumerkel2009} "lambda" library,
supporting only unary and binary application with no free variables.

\begin{prolog-framed}[name=lambdaki]
  :- module(lambdaki, [(\)/2, (\)/3, (^)/3, (^)/4]).

  :- meta_predicate \(1,?), \(2,?,?).
  :- meta_predicate ^(?,0,?), ^(?,1,?,?).

  ~\~(M:Hats, A1)     :- copy_term(Hats, Copy), call(M:Copy, A1).
  ~\~(M:Hats, A1, A2) :- copy_term(Hats, Copy), call(M:Copy, A1, A2).

  ~^~(A1, P, A1)     :- call(P).
  ~^~(A1, P, A1, A2) :- call(P, A2).
\end{prolog-framed}

\subsection{Non-backtrackable mutable state}
\label{app:ccnbstate}

The module "ccnbstate" implements a delimited context "run_nb_state/4"
providing one piece of non-backtrackable mutable state, which can be modified with any binary predicate
using "app/2" or retrieved using "get/2". The global variable used to store the state
is protected from accidental interference and is destroyed when the stateful
computation is finished. 
\begin{prolog-framed}[name=ccnbstate]
  :- module(ccnbstate, [run_nb_state/4, app/2, get/2]).
  :- use_module(library(delimcc)).

  :- meta_predicate app(+,2).
  ~app~(Pr,P) :- p_shift(Pr, app(P)).
  ~get~(Pr,P) :- p_shift(Pr, get(P)).

  :- meta_predicate run_nb_state(+,0,+,-).
  ~run_nb_state~(Pr, Goal, S1, S2) :-
     gensym(nbs,K),
     setup_call_cleanup( nb_setval(K, S1),
                         (run(Pr, Goal, K), nb_getval(K, S2)),
                         nb_delete(K)).

  ~run~(Pr, Goal, K) :- p_reset(Pr, Goal, Status), cont(Status, Pr, K).

  ~cont~(susp(R,Cont), Pr, K) :- handle(R,K), run(Pr, Cont, K).
  ~cont~(done, _, _).

  ~handle~(get(S),K) :- nb_getval(K,S).
  ~handle~(app(P),K) :- nb_getval(K,S1), call(P,S1,S2), nb_setval(K,S2).
\end{prolog-framed}

\subsection{Non-backtrackable mutable references}
\label{app:ccnbref}

This module implements a delimited context "run_nb_ref/1" 
for supplying unique references
to non-backtrackable mutable storage cells using "nbref_new/2".
These cells are guaranteed to be
released when the context is closed. 
\begin{prolog-framed}[name=ccnbref]
  :- module(ccnbref, [run_nb_ref/1, nbref_new/2]).
  :- use_module(library(delimcc), [p_reset/3, p_shift/2]).

  ~nbref_new~(Val, Ref) :- p_shift(nbref, new(Val,Ref)).

  :- meta_predicate run_nb_ref(0).
  ~run_nb_ref~(Goal) :- 
    setup_call_cleanup(setup(E), run(Goal, E), cleanup(E)).

  ~setup~(E) :- gensym(nbref,ID), atom_concat(ID,`.',E), nb_setval(E, 0).
  ~delete~(E,I) :- atomic_concat(E,I,Ref), nb_delete(Ref).
  ~cleanup~(E) :- nb_getval(E, N), nb_delete(E), 
                 forall(between(1,N,I), delete(E,I)).

  ~run~(Goal, E) :- p_reset(nbref, Goal, Status), cont(Status, E).

  ~cont~(done, _).
  ~cont~(susp(new(Val,Ref),Cont), E) :- 
    nb_getval(E, I), J is I+1, atomic_concat(E, J, Ref),
    nb_setval(Ref, Val), nb_setval(E,J).
    run(Cont, E).
\end{prolog-framed}

\subsection{Program transformation}
\label{sec:ccmacros}

The module "ccmacros" implements a shallow program transformation to support
tabling. Predicates decalared `tabled' are renamed (using the special
SWI Prolog hook "prolog:rename_predicate/2" to appendng
a "`#'" to their given name) and the original predicate name defined
as a metacall of the renamed predicate via cctable/1, which is 
assumed to be available in the module where the tabled precicate
is defined. 

\begin{prolog-framed}[name=ccmacros,label=lst:ccmacros]
  :- module(ccmacros, [op(1150, fx, table)]).
  :- op(1150, fx, table).

  ~system:term_expansion~((:- table(Specs)), Clauses) :- 
     foldl_clist(expand_cctab, Specs, Clauses, []).

  ~foldl_clist~(P,(A,B)) --> !, call(P,A), foldl_clist(P,B).
  ~foldl_clist~(P,A)     --> call(P,A).

  ~prolog:rename_predicate~(M:Head, M:Worker) :-
     `&flushed_predicate'(M:`&cctabled'(_)),
     call(M:`&cctabled'(Head)), !,
     head_worker(Head, Worker).

  ~expand_cctab~(Name//Arity) --> !, 
     {A2 is Arity+2}, 
     expand_cctab(Name/A2).

  ~expand_cctab~(Name/Arity) --> 
     { functor(Head, Name, Arity), head_worker(Head, Worker)},
     [ (:- discontiguous(`&cctabled'/1))
     , `&cctabled'(Head)
     , (Head :- cctabled(Worker))
     ]. 

  ~head_worker~(Head, Worker) :-
    Head   =.. [H|As], atom_concat(H, `#', W),
    Worker =.. [W|As].
\end{prolog-framed}

\subsection{Description of library predicates}
\seclab{library}

The module "rbutils" is an interface to the predicates in the standard SWI Prolog
module "library(rbtrees)" but with the argument order of some predicates modified.
The documentation for "rbtrees" is available at
\url{http://www.swi-prolog.org/pldoc/doc/_SWI_/library/rbtrees.pl}.
If we let "rbtree(A,B)" denote the type of trees mapping keys of type "A"
to values of type "B", then the types \citep{SchrijversCostaWielemaker2008}
and meanings of the predicates used in this report are as follows:

{\parindent 0pt
\def\dt{\parskip 0.65em\leftskip 0em}
\def\dd{\parskip 0.15em \par\leftskip 1.5em}

  \dt "~rb_empty~(-rbtree(A,B))" 
  \dd "rb_empty(T)" means "T" is an empty red-black tree.

  \dt "~rb_add~(+A, B, +rbtree(A,B), -rbtree(A,B))"
  \dd "rb_add(K,V,T1,T2)" means "K" is not in "T1" and "T2"  is the result of
      adding value "V" under key "K" to "T1".

  \dt "~rb_in~(A, B, +rbtree(A,B))"
  \dd "rb_in(K,V,T)" means "T" contains value "V" under key "K".

  \dt "~rb_app~(+pred(B,B), +A, +rbtree(A,B), -rbtree(A,B))"
  \dd "rb_app(P,K,T1,T2)" means "T2" is the same as "T1" except that the value
      associated with "K" is "V1" in "T1" and "V2", where "call(P,V1,V2)" is true.

  \dt "~rb_upd~(+A, B, B, +rbtree(A,B), -rbtree(A,B))"
  \dd "rb_app(K,V1,V2,T1,T2)" means "T2" is the same as "T1" except that the
      value associated with "K" is "V1" in "T1" and "V2" in "T2".

  \dt "~rb_fold~(+pred(pair(A,B),S,S), +rbtree(A,B), S, S)"
  \dd "rb_fold(P,T,S1,S2)" means "S2" is the result of folding "P" over a list of key-value
      pairs "K-V" obtained by traversing the tree.

}
\vspace{0.5em}
\noindent
The "store" module, also included in "genutils", uses "rbtrees" to manage an associative map where 
a fresh key is allocated each time a new entry is added. The types and meanings of the main predicates
are:
{\parindent 0pt
\def\dt{\parskip 0.65em\leftskip 0em}
\def\dd{\parskip 0.15em \par\leftskip 1.5em}

  \dt "~store_new~(-store)" 
  \dd {"store_new(S)"} means {"S"} is an empty store

  \dt "~store_add~(A,-ref(A),+store,-store)" 
  \dd "store_add(V,R,S1,S2)" means "S2" is the result of adding value "V" to "S1", and "R" is a new
  reference to the location in "S2" where "V" is stored.

  \dt "~store_app~(+pred(A,A),+ref(A),+store,-store)"
  \dd "store_app(P,R,S1,S2)" means "store_upd(R,V1,V2,S1,S2), call(P,V1,V2)".

  \dt "~store_upd~(+ref(A),A,A,+store,-store)"
  \dd "store_upd(R,V1,V2,S1,S2)" means "S1" and "S2" differ only in the value associated with reference "R",
  which is "V1" is "S1" and "V2" in "S2".

}
\vspace{0.5em}
\noindent
The "tries" library module included with SWI Prolog manages a associative map from arbitrary
terms as keys to mutable storage cells. Key terms which contain variables are considered
to match any term in the same variant class. Value terms are copied both on insertion and
retrieval. Documentation is available at \url{http://www.swi-prolog.org/pldoc/man?section=trie}.


	\bibliographystyle{abbrvnat}
	{\small \bibliography{all,compsci,me}} 
\end{document}